\documentclass[prb,aps,twocolumn,showpacs]{revtex4-1}
\usepackage{amsmath}
\usepackage{amssymb}
\usepackage[greek,english]{babel}
\usepackage{graphicx}
\usepackage{bm}
\newcommand{\ket}[1]{\left|#1\right\rangle}

\newcommand{\expect}[1]{\left\langle #1 \right\rangle}

\newcommand{\sech}{\operatorname{sech}}

\newcommand{\Bcr}{B_{\rm cr}}

\begin{document}
\latintext
\author{Erik Welander}
\thanks{Current affiliation: \textit{Institut f\"{u}r Festk\"{o}rpertheorie, Westf\"{a}lische Wilhelms-Universit\"{a}t, M\"{u}nster, Germany}}
\affiliation{Department of Physics, University of Konstanz, D-78457 Konstanz, Germany}
\author{Julia Hildmann}
\thanks{Current affiliation: \textit{Department of Physics, McGill University, Montreal, Canada}}
\affiliation{Department of Physics, University of Konstanz, D-78457 Konstanz, Germany}
\author{Guido Burkard}
\affiliation{Department of Physics, University of Konstanz, D-78457 Konstanz, Germany}
\title{Influence of Hyperfine Interaction on the Entanglement of Photons Generated by Biexciton Recombination}
\begin{abstract}
The quantum state of the emitted light from the cascade recombination of a biexciton in a quantum dot is theoretically investigated including exciton fine structure splitting (FSS) and electron-nuclear spin hyperfine interactions. In an ideal situation, the emitted photons are entangled in polarization making the biexciton recombination process a
candidate source of entangled photons necessary for the growing field of quantum communication and computation. The coherence of the exciton states in real quantum dots is affected by a finite FSS and the hyperfine interactions via the effective magnetic field known as the Overhauser field. We investigate the influence of both sources of decoherence and find that although the FSS combined with a stochastic exciton lifetime is responsible for the main loss of entanglement, the two effects cannot be minimized independently of each other. Furthermore, we examine the possibility of reducing the decoherence from the Overhauser field by partially polarizing the nuclear spins and applying an external magnetic field. We find that an increase in entanglement depends on the degree as well as the direction of the nuclear spin polarization.  
\end{abstract}
%
%
%
%
%
%
%
%
%
\maketitle
\section{Introduction}
A reliable source of entangled photons is a requirement for many protocols used in the rapidly developing field of quantum communication\cite{RevModPhys.74.145}. An established method for creating polarization entangled photons is
by parametric down-conversion\cite{PhysRevLett.75.4337,PhysRevA.60.R773}. However, this technique suffers
from being both inefficient and stochastic, causing problems since many quantum communication protocols require
an on-demand source. An alternative source is thus desired, and here we consider the biexciton cascade
recombination\cite{benson,PhysRevLett.108.040503}. In a quantum dot (QD), the biexciton, which is composed of two conduction band electrons and two valence band holes, can under ideal
conditions recombine under the emission of two photons entangled in polarization\cite{benson,santori}.
The biexciton recombines via one of two possible intermediate exciton states, each consisting of one
conduction band electron and one valence band hole.

In most quantum dots, the two optically active
exciton states are energetically separated by a quantity known as the fine structure splitting (FSS) arising from higher order electron-hole exchange interactions\cite{bayer1,PhysRevB.81.045311,welander2}. 
A finite FSS can affect the coherence of the emitted two-photon state in two ways. If the FSS
is larger
than the linewidth of the emitted light, the photons become distinguishable via a simple frequency
measurement which reveals the ``which-way'' information and destroys the entanglement\cite{santori}.
However, even if the splitting is smaller than the linewidth but still finite,
the initially coherent exciton 
state acquires a random phase before recombining, due to the stochastic life time.
Methods of reducing and eliminating
the FSS include applying magnetic\cite{stevenson} and electric\cite{gerardot,bennett,bennett2,ghali,vogel,hogele,welander2}
fields as well as strain\cite{plumhof,trotta,seidl,PhysRevB.88.155330}.

In addition to the dephasing from a finite FSS,
the intermediate exciton state is affected by the spins of the $10^5$-$10^6$ nuclei present in a III-V group
semiconductor QD. The spins of the
electron\cite{PhysRevB.65.205309,PhysRevLett.88.186802,Koppens2006,petta} and
hole\cite{PhysRevB.78.155329,PhysRevB.79.195440,PhysRevLett.102.146601} constituting the exciton couple to nuclear
spins via the hyperfine interaction
and are subject to an effective nuclear magnetic field, known as the Overhauser field.
Because of the large number of nuclear spins, the Overhauser field can be considered to be
stochastic and is another source of decoherence, including an additional random
phase of the intermediate exciton state.

Experimentally, various techniques for creating entangled light using semiconductor
microstructures have been demonstrated\cite{PhysRevLett.108.040503, Dousse2010, stevenson, zwiller1,1405.3765}. One successful approach relies on the application of an external magnetic field perpendicular to the growth direction of the QD\cite{stevenson,stevenson1}, which tunes the energy levels of the optically active excitons by hybridization to optically inactive states. This requires the in-plane $g$-factors of the electrons and holes to have opposite signs. InAs dots with AlGaAs barrier material having this property were reported\cite{stevenson1} and by applying an in-plane magnetic field, the FSS were tuned to zero. Nevertheless, a complete theoretical explanation of the partial loss of entanglement is still missing. Understanding the dynamics of the intermediate exciton states is essential to investigate the entanglement of the emitted light. In fact, any dephasing and loss of coherence of the exciton state will be reflected in the final photon state.

In this paper, we investigate the interplay between the dephasing due to a finite FSS together with a stochastic recombination time and the decoherence caused by the hyperfine interaction. Our results concern a dot for which the FSS is tuned to zero by an in-plane magnetic field, which requires the in-plane electron and hole $g$-factors to have opposite signs. One way of reducing the fluctuations of the Overhauser field is by dynamic nuclear spin polarization\cite{PhysRevB.70.195340,PhysRevLett.102.216802}, causing the nuclear spins to have a preferred direction. The polarized nuclear ensemble produces a finite effective magnetic field, which may modify the exciton energies and eigenstates. Since the elimination of the FSS has been demonstrated using an external in-plane magnetic field, a tempting idea could be to use the effective magnetic field produced by the polarized nuclear ensemble, to reduce both the FSS and the Overhauser field fluctuations. However, this turns out not to be possible, since an effective in-plane magnetic field for both holes and electrons is required. For heavy holes, the in-plane component of the nuclear magnetic field vanishes. Furthermore, we show that the effect from the two sources of decoherence cannot be minimized independently of each other. 

We consider the effect of finite nuclear spin polarization and find that the entanglement of the emitted light can be improved by nuclear spin polarization. The efficiency of the entanglement improvement depends on the direction along which the nuclear spins are polarized, which is explained by the fact that the hyperfine coupling tensor is not isotropic. We find that the maximum enhancement of the entanglement is achieved when nuclear spins are polarized along the direction for which the coupling tensor has its largest components, in our case in the growth direction of the quantum dot. 

A nuclear polarization in the growth direction gives an additional contribution to the FSS and therefore increases the dephasing. To cancel the effective nuclear field in the growth direction and minimize the FSS at the same time, we propose applying an external magnetic field along a specific direction having an in-plane and a perpendicular component. Combining a finite nuclear spin polarization along the growth direction of the QD with an external magnetic field, we find a significant improvement of the two-photon state entanglement.

\begin{figure}
\includegraphics{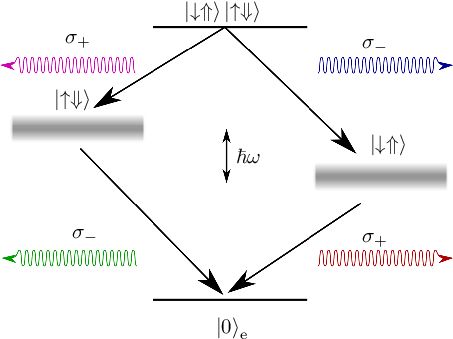}
\caption{\label{cascade}Energy diagram of the exciton states in the quantum dot. The topmost level is the biexciton state $\ket{\downarrow\Uparrow}\ket{\uparrow\Downarrow}$ consisting of two excitons. The two intermediate levels are the exciton states $\ket{\downarrow\Uparrow}$ and $\ket{\uparrow\Downarrow}$ energetically separated by the FSS $\hbar\omega$, which depends on the built-in FSS $|\delta_1|$ and an applied magnetic field. The lowest level is the semiconductor ground state $\ket{0}_{\rm e}$ containing no excitons. The biexciton recombines to one of the intermediate exciton states under the emission of a photon with polarization depending on the exciton state. The remaining exciton then recombines under the emission of another photon of orthogonal polarization. In the ideal situation, this creates a pair of photons entangled only in polarization. The existence of a FSS can degrade the entanglement, even if $\hbar\omega < \Gamma$ where $\Gamma$ is the linewidth of the emitted light because of the stochastic lifetime of the exciton which causes dephasing and leads to a statistical mixture instead of a pure state. Furthermore, because of the interaction with stochastic magnetic field originating from the nuclear spins in the quantum dot, the exciton energy levels are not sharply defined, meaning that $\omega$ also fluctuates which causes even more dephasing.}
\end{figure}
\section{Theoretical model}
We consider a QD of a cubic semiconductor containing one exciton consisting of one electron and one heavy hole. The spins of the electron and the hole couple to the spins of the atomic nuclei in the QD. This effect is to a good approximation described by the contact hyperfine Hamiltonian
\begin{equation}
\label{hyperfine}
H_{\rm HF} = \sum_{n = 1}^N\mathbf I^{(n)} \cdot \left(\mathbf A_{\rm e}^{(n)} \mathbf S^{\rm e} + \mathbf A_{\rm h}^{(n)} \mathbf S^{\rm h}\right),
\end{equation}
where the summation runs over all nuclear spins in the dot, $\mathbf I^{(n)}$ is the spin operator of the $n$-th nuclear spin, $\mathbf S^{\rm e(h)}$ is the electron(hole) spin operator, and $\mathbf A_{\rm e(h)}^{(n)}$ are the hyperfine coupling tensors between the $n$-th nuclear spin and corresponding electron(hole) spin component. In this work, we are interested in heavy holes, for which only the $z-z$ component of the $\mathbf A_{\rm h}^{(n)}$ tensor is finite. This allows us to define an effective magnetic field by moving the electron and hole spin operators outside of the summation and obtain
\begin{equation}
H_{\rm HF} = \mu_B\mathbf B_{\rm HF}\cdot \mathbf f_{\rm e} \mathbf S^{\rm e}  +  \mu_B B^z_{\rm HF}f_{\rm h}S^{\rm h}_z,
\end{equation}
where $\mathbf f_{\rm e}$ is the electron-nuclear spin coupling tensor. The heavy hole - nuclear spin coupling is of Ising type, described by the coupling constant  $f_h$ between the $z$-component of hole and nuclear spins. The vector $\mathbf B_{\rm HF}$ is known as the Overhauser field and acts like an effective magnetic field from the perspective of the exciton. Because of the large number of nuclear spins\cite{petta} ($10^5 - 10^6$) for a typical quantum dot, the Overhauser field is often modelled as a stochastic magnetic field and will be considered further in Section \ref{polarsect}. We consider the case of diagonal hyperfine coupling tensors, which for an electron in a III-V semiconductor QD can be written\cite{coish} as $A_{x,y,z}^{(n)} \propto g_{\rm e}^{x,y,z}|F(\mathbf r_n)|^2$, where $g_{\rm e}^{x,y,z}$ are the electron $g$-factors and $|F(\mathbf r_n)|^2$ is the electron density at the atomic site $n$. An important feature is the dependence of the spatial direction $x,y,z$, which indicates that fluctuations of the different spatial components of the Overhauser field influence the energy of the electron differently.

A basis for the Hilbert space of the electron spin $\mathcal H_e$ is given by $\{\ket{\uparrow},\ket{\downarrow}\}$, where $\uparrow(\downarrow)$ corresponds to the spin $s_z = 1/2$ ($s_z = -1/2$) state, and for the heavy hole the Hilbert space $\mathcal H_h$ is spanned by $\{\ket{\Uparrow},\ket{\Downarrow}\}$, with $\Uparrow(\Downarrow)$ corresponding to the hole spin states $j_z = 3/2$ ($j_z=-3/2$). The Hilbert space of the exciton is given by the product space $\mathcal H_X = \mathcal H_e \otimes \mathcal H_h$ and is spanned by the basis vectors $\{\ket{\downarrow\Uparrow}, \ket{\uparrow\Downarrow},\ket{\uparrow\Uparrow},\ket{\downarrow\Downarrow}\}$. The states $\ket{\uparrow\Downarrow}$ and $\ket{\downarrow\Uparrow}$ are known as bright since they can recombine under the emission of a single photon whereas $\ket{\uparrow\Uparrow}$ and $\ket{\downarrow\Downarrow}$ are known as dark. The idealized recombination chain of the biexciton is given by
\begin{gather*}
\ket{0}_{\rm ph}\otimes\ket{\downarrow\Uparrow}\ket{\uparrow\Downarrow}\\
\downarrow\\
\frac{\ket{\sigma_+}\otimes\ket{\uparrow\Downarrow} + \ket{\sigma_-}\otimes\ket{\downarrow\Uparrow}}{\sqrt{2}} \\
\downarrow\\
\frac{\ket{\sigma_+}\ket{\sigma_-}+\ket{\sigma_-}{\ket{\sigma_+}}}{\sqrt 2}\otimes\ket{0}_{\rm e}
\end{gather*}
where $\ket{\sigma_\pm}$ are photon states of circularly polarized light, and $\ket{0}_{\rm ph(e)}$ is the photon (crystal) vacuum. In reality the intermediate state undergoes time-evolution before the exciton has recombined which may lead to degradation of the entanglement of the emitted light. The final state of the intermediate exciton state can be written using a density matrix
\begin{equation}
\label{excirho}
\boldsymbol\rho_X = \begin{pmatrix}p & \gamma\\
\gamma^\ast & 1-p\end{pmatrix},
\end{equation}
where $p$ and $1 - p$ are the populations of the states $\ket{\sigma_+}\otimes\ket{\uparrow\Downarrow}$ and $\ket{\sigma_-}\otimes\ket{\downarrow\Uparrow}$, and $\gamma$ is the off-diagonal matrix element required to describe a quantum mechanical superposition of the basis states. Here, only there bright excitons are taken into consideration. The concurrence of the emitted light is then given by\cite{wootters}
\begin{equation}
\label{conceq}
C = 4\sqrt{p(1-p)}|\gamma|,
\end{equation}
where we note that the off-diagonal elements are essential for the entanglement.

Under the influence of a magnetic field $\mathbf B = (B_x, B_y, B_z)^T$ the exciton system is described by the Hamiltonian\cite{bayer1}
\begin{equation}
\label{origham}
H = \frac{1}{2}\begin{pmatrix} \delta_0 - h_{z-} & \delta_1 & h_e & h_h \\
\delta_1^\ast & \delta_0 + h_{z-} & h_h^\ast & h_e \\
h_e^\ast & h_h & -\delta_0 + h_{z+} & \delta_2 \\
h_h^\ast & h_e^\ast & \delta_2^\ast & -\delta_0 -h_{z+}\end{pmatrix},
\end{equation}
in the heavy exciton basis $\{\ket{\downarrow\Uparrow},\ket{\uparrow\Downarrow},\ket{\uparrow\Uparrow},\ket{\downarrow\Downarrow}\}$, where $\delta_0$ is the splitting between bright and dark excitons, $|\delta_{1(2)}|$ is the FSS for bright (dark) excitons, $h_{z\pm} = \mu_BB_z(g^z_e \pm g^z_h)$, $h_{e(h)} = \mu_B\left(B_xg^x_{e(h)} +iB_yg^y_{e(h)}\right)$ and $g^\alpha_{e(h)}$ are effective $g$-factors for electrons(holes) along the $\alpha$-axis.
For the case $\mathbf B = B_x \hat x$ and $\delta_1, \delta_2 \in \mathbb{R}$, the eigenvalue problem $H\ket{X} = E\ket{X}$ is analytically solvable and the two eigenenergies corresponding to the two bright excitons are given by
\begin{subequations}
\begin{align}
E_1 &= \frac{\delta_1 + \delta_2+ \sqrt{(2\delta_0 + \delta_1 - \delta_2)^2 + 4(g_e^x + g_h^x)^2\mu_{\rm B}^2B_{x}^2} }{4}\\
E_2 &= \frac{-\delta_1 - \delta_2 +\sqrt{(2\delta_0 - \delta_1 + \delta_2)^2 + 4(g_e^x - g_h^x)^2\mu_{\rm B}^2B_{x}^2}}{4}.
\end{align}
\end{subequations}
Similar expressions were found by Bayer \textit{et. al.}\cite{PhysRevB.61.7273}, where it should be noted that our expressions differ slightly, due to a sign error. Demanding the bright exciton states to be degenerate, i.e. $E_1 = E_2$, gives an equation for a critical field $B_{\rm cr}$,
\begin{align}
\label{bcreq}
\begin{split}
&2(\delta_1 + \delta_2) + \sqrt{(2\delta_0 + \delta_1 - \delta_2)^2 + 4(g_e^x + g_g^x)^2\mu_{\rm B}^2B_{\rm cr}^2} \\
= &\sqrt{(2\delta_0 - \delta_1 + \delta_2)^2 + 4(g_e^x - g_g^x)^2\mu_{\rm B}^2B_{\rm cr}^2},
\end{split}
\end{align} 
for which the FSS vanishes. The FSS can be written $E_1 - E_2$, which may be expanded in Maclaurin series in $B_x$ to the second order, which gives the approximation
\begin{equation}
\hbar\omega(B_x) \approx \delta_1 + \frac{\mu_{\rm B}^2\left[4g_e^xg_h^x\delta_0 - \left({g_e^x}^2 + {g_h^x}^2\right)(\delta_1 - \delta_2)\right]}{4\delta_0^2 - (\delta_1 - \delta_2)^2}B_x^2.
\end{equation}
A similar expression was also presented by Stevenson \textit{et al.}\cite{stevenson1}. Solving $\omega(B_{\rm cr}) = 0$ provides an expression for the critical magnetic field, given by
\begin{equation}
B_{\rm cr} = \pm \sqrt{\frac{\delta_1\left[(\delta_1 - \delta_2)^2 - 4\delta_0^2\right]}{\mu_{\rm B}^2\left[4g_e^xg_h^x\delta_0 - \left({g_e^x}^2 + {g_h^x}^2\right)(\delta_1 - \delta_2)\right]}}.
\end{equation}

For the general case involving arbitrary magnetic fields and complex $\delta_1,\delta_2$, an analytical diagonalization of the Hamiltonian in Eq. (\ref{origham}) is not known. However, the energy splitting between the bright and dark excitons $\delta_0$ is larger than all other relevant energies, including the magnetic coupling elements. Therefore, we can apply the Schrieffer-Wolff\cite{PhysRev.149.491,winkler} transformation, which provides us with an effective Hamiltonian for the bright exciton subspace:
\begin{equation}
\tilde H = \tilde H_0 + \tilde H_2,
\end{equation}
where
\begin{equation}
\label{SW1}
\tilde H_0 = \frac{1}{2}\begin{pmatrix} \delta_0 - h_{z-} & \delta_1  \\
\delta_1^\ast & \delta_0 + h_{z-}
\end{pmatrix},
\end{equation}
and
\begin{equation}
\label{SW2}
\tilde H_2 = \frac{1}{4}\begin{pmatrix}
\frac{|h_e|^2}{\delta_0 - b_{e}} + \frac{|h_h|^2}{\delta_0 + b_{h}} & \frac{\delta_0 h_eh^\ast_h}{\delta_0^2 - b_e^2} + \frac{\delta_0 h_eh^\ast_h}{\delta_0^2 - b_h^2}\\
\frac{\delta_0 h^\ast_eh_h}{\delta_0^2 - b_e^2} + \frac{\delta_0 h^\ast_eh_h}{\delta_0^2 - b_h^2} & \frac{|h_e|^2}{\delta_0 + b_{e}} + \frac{|h_h|^2}{\delta_0 - b_{h}}
\end{pmatrix},
\end{equation}
with $b_{e(h)} = \mu_BB_zg^z_{e(h)}$. The FSS $\hbar\omega$ can now be found by solving the eigenvalue problem
\begin{equation}
\tilde H\ket{n} = E_n\ket{n}
\end{equation}
for $n = 1,2$ and taking the difference $\hbar\omega = |E_1 - E_2|$. Using the explicit form of $\tilde H$ given by Eqs. (\ref{SW1}) and (\ref{SW2}) we find 
\begin{equation}
\label{FSS1}
\hbar\omega = \sqrt{\Omega_1^2 + \Omega_2^2},
\end{equation}
where 
\begin{align}
\label{FSS2}
\Omega_1 &= h_{z-} - \frac{1}{2}\frac{|h_e|^2b_e}{\delta_0^2 - b_{e}^2} + \frac{1}{2}\frac{|h_h|^2b_h}{\delta_0^2 - b_{h}^2},\\
\label{FSS3}
\Omega_2 &=\left|\delta_1 + \frac{1}{4}\frac{\delta_0 h^\ast_eh_h}{\delta_0^2 - b_e^2} + \frac{1}{4}\frac{\delta_0 h^\ast_eh_h}{\delta_0^2 - b_h^2}\right|.
\end{align}
Demanding that $\omega(\mathbf B_{\rm cr}) = 0$ again defines critical magnetic fields, not necessarily along $\hat x$, for which the FSS is eliminated. Inspecting Eq. (\ref{FSS1}), we realize that $\omega = 0$ only if $\Omega_1 = 0$ and $\Omega_2 = 0$. $\Omega_1$ can be tuned to zero by adjusting the total magnetic field along $\hat z$. $\Omega_2$ depends on in in-plane components of the magnetic field, $B_x$ and $B_y$ and the bright exciton coupling $\delta_1$, which may be complex. The phase of $\delta_1$ is related to the geometry of the quantum dot whereas the phase of $h_e^\ast h_h$ depends on the anisotropy of the $g$-tensors. For a quantum dot with an isotropic in-plane $g$-tensor, $h_e^\ast h_h$ is real and in this situation, no magnetic field can completely eliminate the FSS caused by a complex $\delta_1$. However, $\delta_1\in\mathbb R$ combined with a isotropic $g$-tensor, Eq. (\ref{FSS3}) reveals that $h_e^\ast h_n< 0$ is a criterion for the existence of a magnetic field such that $\Omega_2 = 0$. In turn, this requires that the in-plane $g$-factors of the electron and hole have opposite signs and this implies that not all quantum dots can be tuned to support degenerate bright excitons, also noted in experimental work\cite{stevenson1,PhysRevB.61.7273} where InAs dots surrounded by different barrier materials were studied. The corresponding in-plane $g$-factors were extracted, and here the values for the case of Al$_{0.33}$Ga$_{0.67}$As\cite{stevenson1} as the barrier material are given in Table \ref{paramstbl}. Experimental\cite{PhysRevB.76.041301,PhysRevB.87.161302,PhysRevB.92.165307} and theoretical\cite{PhysRevB.89.115438} studies show that the $g$-factor of InAs and InGaAs QDs can be tuned over large range of values and here we use $g_e^z = 2$, $g_h^z = -5$ to give a total exciton $g$-factor $g_e^z + g_h^z = -3$, measured in experiments\cite{PhysRevB.76.041301}.

There are two main sources of loss of entanglement: (1) the fine-structure splitting combined with the stochastic exciton life time and (2) the stochastic Overhauser field that affect the intermediate exciton state. Both mechanisms lead to the acquisition of an unknown phase which causes a reduction of the entanglement. To investigate further, we consider the effect of the two above-mentioned mechanisms on the density operator of the intermediate exciton state. We choose the diagonal basis $\ket{1},\ket{2}$ which are eigenvectors of $\tilde H$, and an initial density operator in matrix form as
\begin{equation}
\boldsymbol\rho(t=0) = \begin{pmatrix}\rho_{11} & \rho_{12} \\ \rho_{21} & \rho_{22}
\end{pmatrix},
\end{equation}
where $\rho_{11} + \rho_{22} = 1$ and $\rho_{12}^\ast = \rho_{21}$. We can now determine the time-evolution for the density operator via the Heisenberg equation of motion,
\begin{equation}
i\hbar\dot{\boldsymbol\rho} = [\boldsymbol{\rho},\tilde H],
\end{equation}
which has the solution
\begin{equation}
\boldsymbol\rho(t > 0) = \begin{pmatrix}\rho_{11} & \rho_{12}e^{i\omega t} \\ \rho_{21}e^{-i\omega t} & \rho_{22}
\end{pmatrix},
\end{equation}
with the FSS $\omega$ given by Eqs. (\ref{FSS1})--(\ref{FSS3}). If the FSS is stochastic as one would expect from an Overhauser field we may find its contribution by statistical averaging
\begin{equation}
\expect{\boldsymbol\rho} = \int_{-\infty}^\infty f_\Omega(\omega)\begin{pmatrix}\rho_{11} & \rho_{12}e^{i\omega t} \\ \rho_{21}e^{-i\omega t} & \rho_{22}\end{pmatrix} d\omega,
\end{equation}
where $f_\Omega(\omega)$ is the probability density function of the FSS.
\begin{figure}
\centering
\includegraphics{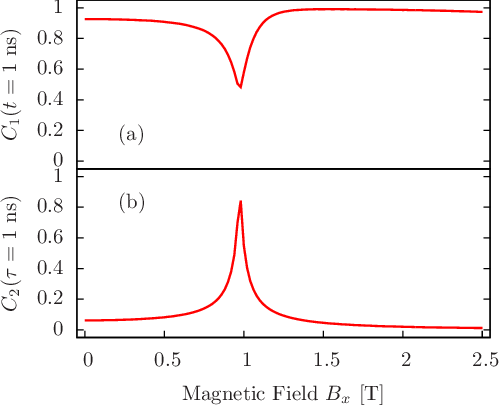}
\caption{\label{BfieldFig}
Comparison between the concurrence of a stochastic FSS with a definite recombination time $C_1(t)$ and the concurrence in the presence of a known FSS with a stochastic recombination time $C_2(\tau)$. (a) The concurrence of the entangled light $C_1(t)$, when a stochastic Overhauser field with $\sigma_x = \sigma_y = \sigma_z = 5$ mT causes an unknown relative phase between the exciton states when the exciton undergoes time-evolution of $t = 1$ ns. (b) The concurrence of the entangled light $C_2(\tau)$ when the known FSS causes an unknown relative phase between the exciton states. The exciton undergoes time-evolution for a stochastic recombination time, which on average is given by $\tau = 1$ ns. We see that there is a trade-off when attempting to minimizing both sources of decoherence since at $B_x = \Bcr \approx 1$ T, $C_2(\tau)$ has a maximum, while $C_1(t)$ has a minimum. This indicates that both sources of decoherence have to be taken into consideration simultaneously.}
\end{figure}
For the quantum dot hosting the exciton we assume a stochastic Overhauser field with Gaussian distribution. The FSS, however, does not have a Gaussian distribution because of the nonlinear way the exciton eigenenergies depend on an applied magnetic field, given by Eqs. (\ref{FSS1})--(\ref{FSS3}). Therefore, the statistical averaging is performed by considering a Gaussian distribution for the Overhauser field with the probability density function
\begin{equation}
\label{gaussdist}
f_{\mathbf B}(\mathbf B) =  \frac{1}{\sigma_x\sigma_y\sigma_z(2\pi)^{3/2}}e^{-B_x^2/2\sigma_x^2 - B_y^2/2\sigma_y^2 - B_z^2/2\sigma_z^2},
\end{equation}
where $\sigma_x$, $\sigma_y$, $\sigma_z$ are the standard deviations of the Overhauser field along $\hat x$, $\hat y$, $\hat z$. We numerically evaluate
\begin{equation}
\label{numer1}
\expect{\boldsymbol\rho} = \int_{-\infty}^{\infty}f_{\mathbf B}(\mathbf B)\begin{pmatrix}\rho_{11} & \rho_{12}e^{i\omega(\mathbf B)t} \\ \rho_{21}e^{-i\omega(\mathbf B)t} & \rho_{22}\end{pmatrix} d\omega,
\end{equation}
using the parameter values given in Table \ref{paramstbl}, from which we can extract an entanglement measure, here the concurrence $C_1(t)$ by using Eq. (\ref{conceq}). An external magnetic field, $\mathbf B^0$ may also be present, and both the Overhauser field and the external magnetic field are included by replacing $g_{e(h)}^\alpha B_\alpha \longrightarrow g_{e(h)}^\alpha B^0_\alpha + f_{e(h)}^\alpha B_\alpha$ in Eqs. (\ref{origham})-(\ref{FSS3}), where $\alpha \in \{x,y,z\}$ and $f_{(h)}^x = f_{(h)}^y = 0$.

To investigate the effect of the stochastic exciton life time we consider a Poissonian recombination process which corresponds to an exponential life time $t$ with probability density function $f_t(t) = e^{-t/\tau}/\tau$ where $\tau$ is the average life time. Calculating the statistical average of the density matrix gives
\begin{align}
\label{concrnd}
\begin{split}
\expect{\boldsymbol\rho} &= \int_{0}^\infty\frac{e^{-t/\tau}}{\tau}\begin{pmatrix}\rho_{11} & \rho_{12}e^{i\omega t} \\ \rho_{21}e^{-i\omega t} & \rho_{22}\end{pmatrix} dt\\
&=\begin{pmatrix}\rho_{11} & \frac{\rho_{12}}{1 + i\omega \tau} \\ \frac{\rho_{21}}{1 - i\omega\tau} & \rho_{22} \end{pmatrix},
\end{split}
\end{align}
which also have decaying concurrence\cite{hohenester} $C_2(\tau) \propto |1~+~i~\omega~\tau~|^{-1}$ for $\omega \neq 0$. This suggests choosing $\omega \ll 1/\tau$ to maximize the concurrence. The two different concurrences are shown in Fig. \ref{BfieldFig}.
We can see that there is a target conflict when applying a magnetic field along $\hat x$. For a critical magnetic field strength $B_x = \Bcr$ the fine structure splitting is eliminated and $C_2(\tau)$ has a maximum, but the concurrence $C_1(t)$ when considering a stochastic magnetic field from the nuclear spins has a minimum. The reason is that the FSS is most sensitive to changes in the magnetic field at this point. To obtain a more complete picture we need to take both sources of decoherence into account simultaneously, which we achieve by averaging the concurrence in Eq. (\ref{concrnd}) using the probability distribution for the stochastic magnetic field $f_{\mathbf B}(\mathbf B)$ given by Eq. (\ref{gaussdist}), which is done numerically by evaluating
\begin{equation}
C(\tau) = \left|\int_{-\infty}^\infty\frac{f_\mathbf{B}(\mathbf B)}{1 + i\omega(\mathbf B) \tau}\,d\omega\right|.
\end{equation}
\section{Results}
\begin{figure}[b]
\includegraphics{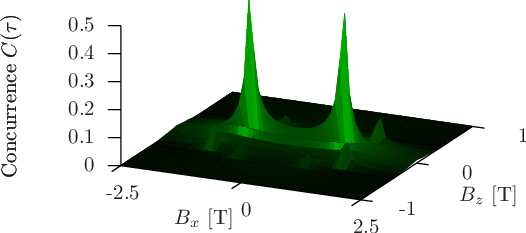}
\caption{\label{totConcFig}Concurrence when the combined effect of dephasing from a stochastic magnetic field due to the nuclear spins and a stochastic exciton lifetime is considered. The maxima occur when the FSS is eliminated and reach values limited by the dephasing from the stochastic Overhauser field. This indicates that the dominant source of decoherence is the FSS combined with the stochastic lifetime of the exciton. An improvement of the maximum concurrence can be achieved by reducing the Overhauser field fluctuations. There are also four local maxima located roughly at $|B_x| = 1.5$ T, $|B_z| = 0.25$ T, demonstrating the importance of treating both sources of decoherence simultaneously.}
\end{figure}
To obtain quantitative results, we choose a set of parameter for the quantum dot given in Table \ref{paramstbl}. 
\begin{table}
\begin{tabular}{|c|c|c|c|c|c|c|}
\hline
$g_e^{x,y}$ & $g_h^{x,y}$  & $g_e^z$ & $g_h^z$ & $\delta_0$ & $\delta_1$ & $\tau$\\
\hline 
$1.21$ & $-0.13$ & $2$ & $-5$ & $50$ {\greektext m}eV & $10$ {\greektext m}eV & 1 ns\\
\hline
\end{tabular}
\caption{\label{paramstbl}Table of the parameters for an InAs quantum dot surrounded by GaAs, used in the numerical calculations, reported from experiments\cite{stevenson1}. The condition that $g_e^xg_h^y < 0$ is necessary to allow the elimination of the FSS. $\delta_1$ is in general complex, but here chosen real. It should be noted, that these parameters are not representative for all quantum dots, but selected in order to allow the FSS to be tuned to zero.}
\end{table}
\subsection{Dominant Source of Decoherence}
In order to improve the concurrence we first establish which source of decoherence causes more loss of concurrence, the FSS or the Overhauser field. From Fig. \ref{BfieldFig} this is not obvious, because at $B_x = \Bcr$ the FSS is minimized but the dephasing from the Overhauser field is maximized. Taking both into account and allowing in addition a magnetic field to be applied along $\hat z$ as well we find the concurrence as function of the applied magnetic field depicted in Fig. \ref{totConcFig}. From our calculations we find a maximum value for the concurrence $C_{\rm max} \approx 0.54$. This is consistent with experimentally reported values of typically around $0.5$\cite{1405.3765,young,hafenbrak}. However, it should be noted, that a broad range of values for the concurrence have been reported, spanning between $0.16$\cite{muller} and $0.8$\cite{zwiller1}.

We see that the two global maxima are located at $(B_x = \pm \Bcr, B_y = 0, B_z = 0)$ which indicates that the FSS is a stronger source of decoherence than the Overhauser field. Still, the concurrence does not reach unity but is rather close to the minimum observed in Fig. \ref{BfieldFig}a. From these observations we conclude that in order to maximize the concurrence, we should keep $B_x = \pm\Bcr$ to eliminate the FSS and now focus on reducing the uncertainty of the Overhauser field. One way of achieving this is to polarize the nuclear spins, which has been experimentally realized\cite{PhysRevLett.96.167403,PhysRevLett.94.047402,PhysRevLett.104.066804,Munsch2014,PhysRevLett.98.107401,PhysRevB.80.035326,PhysRevB.74.245306}, and is investigated in the next section. In addition to the two global maxima, there are four local maxima located close to $|B_x| = 1.5$ T, $|B_z| = 0.25$ T. Although the concurrence is smaller at these points than at the global maxima, they indicate the significance of including the effects of both sources of decoherence simultaneously. 
\subsection{Effect of Nuclear Spin Polarization}
\label{polarsect}
\begin{figure}[t]
\includegraphics{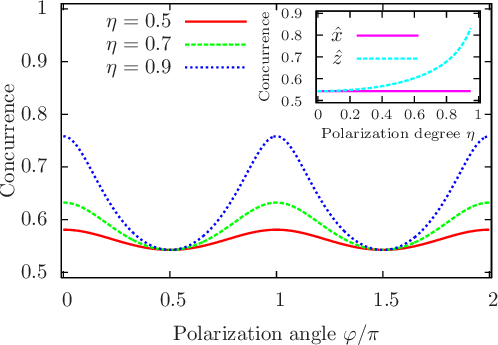}
\caption{\label{anglefig}The concurrence under the condition $(B_x, B_y, B_z) = (\Bcr, 0, 0)$ for different degrees $\eta$ and angles $\varphi$ of the nuclear spin polarization. The nuclear spins are polarized along $(\sin\varphi, 0, \cos\varphi)$ and we can observe a strong dependence on the angle. For a polarization along $\hat x$ ($\varphi = \pm\pi/2$) the increase of concurrence is almost absent in comparison to a polarization along $\hat z$ ($\varphi = n\pi$, $n \in \mathbb{Z}$). Inset: Concurrence as function of the nuclear spin polarization along $\hat x$ (purple) and $\hat z$ (cyan). A nuclear polarization along $\hat z$ leads to a significant improvement in concurrence whereas this effect is all but absent in the case of polarization along $\hat x$. Although polarization along any direction $\mathbf n$ would lead to a reduction of the fluctuations of the Overhauser field along $\mathbf n$, because the in-plane $g$-factors are smaller than along the growth direction, in-plane fluctuations have less effect on the FSS.}
\end{figure}

It is clear that, within our model, when the FSS is eliminated, the remaining reduction of the entanglement originates from the Overhauser field. To investigate how the fluctuations of the Overhauser field vary as function of the nuclear spin polarization we consider a simple model for the Overhauser field along one spatial direction
\begin{equation}
B = \sum_{n = 1}^MA_n\beta_n,
\end{equation}
where $M$ is the number of nuclear spins, $\beta_n$ are binary stochastic variables taking the values $\pm 1$ with probability
\begin{equation}
p_\beta(\pm 1) = \frac{e^{\pm S}}{e^{S} + e^{-S}},
\end{equation}
where
\begin{equation}
S = \frac{\mu_{\rm B}g_{\rm N}B_0}{k_{\rm B}T_{\rm N}},
\end{equation}
$g_N$ is the nuclear $g$-factor, $B_0$ is an external magnetic field and $T_N$ is the nuclear spin temperature.
\begin{figure}[t!]
\includegraphics{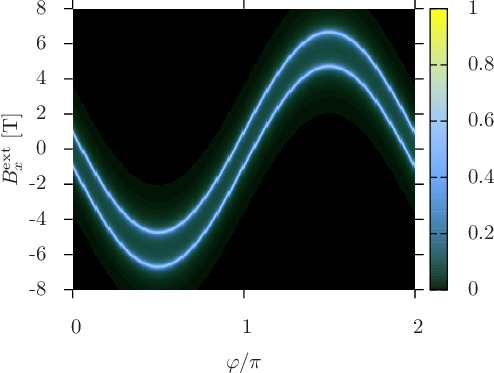}
\caption{\label{concurrencePol1}Concurrence as function of nuclear polarization angle $\varphi$ and applied magnetic field $B_x^{\rm ext}$ along $\hat x$. We assume a fixed degree of $\eta = 90\%$ polarization along $(n_x,0,n_z) = (\sin\varphi, 0, \cos\varphi$). The total magnetic field is given by $\mathbf B = (B_x^{\rm ext} + \eta B_{\rm max}\sin\varphi, 0, 0$), which implies an applied magnetic field $B_z^{\rm ext} = - \eta B_{\rm max}\cos\varphi$ along $\hat z$. For every angle of polarization, there are two values of the external magnetic field maximizing the concurrence, which are located at $B_x = \pm\Bcr$, which minimizes the FSS.}
\end{figure}
The polarization $\eta$ is given by
\begin{equation}
\eta = \expect{\beta} = \frac{e^{S} - e^{-S}}{e^{S} + e^{-S}} = \tanh S,
\end{equation}
and the variance is consequently
\begin{equation}
\sigma^2 = \expect{\beta^2} - \expect{\beta}^2 = \sech^2 S = 1-\eta^2.
\end{equation}
In Appendix \ref{weightedaverage} it is shown that 
\begin{equation}
\label{normallimit}
B \sim \mbox{N}(M \eta, C(1 - \eta^2)), \mbox{ when } M \rightarrow \infty,
\end{equation}
where $\mbox{N}(\mu, \sigma^2)$ is a Gaussian distribution with mean $\mu$ and standard deviation $\sigma$, $M$ is the number of nuclear spins, and $C$ depends on $M$ and $A_n$. Typically, $C$ will have to be determined experimentally or by numerical simulations and we do not attempt to calculate it here, but the general form Eq. (\ref{normallimit}) does not depend on the specific QD. 
Since the fluctuations of the Overhauser field decrease with increasing polarization we now assume that the nuclear spins are polarized to degree $\eta$ along $\mathbf n = (n_x, n_y, n_z)^T$, where $n_x^2 + n_y^2 + n_z^2 = 1$. The assumption that the nuclear spin can be polarized along an arbitrary direction relies on experimental demonstrations\cite{Makhonin2011}. This gives an effective magnetic field $\mathbf B_{\rm HF} = B_{\rm max}\eta\mathbf n$, with variances
\begin{equation}
(\sigma_x^2, \sigma_y^2, \sigma_z^2) = \left(C_x[1-\eta^2n_x^2], C_y[1-\eta^2n_y^2],C_z[1-\eta^2n_z^2)]\right).
\end{equation}
Together with the applied magnetic field $\mathbf B_{\rm ext} = (B_x^{\rm ext}, B_y^{\rm ext}, B_z^{\rm ext})^T$ the total effective magnetic field depends on 7 variables: $B_x^{\rm ext}, B_y^{\rm ext}, B_z^{\rm ext}, n_x, n_y, n_z,$ and $\eta$. In order to narrow the search for optimal parameters, we make the following observations: first, $\hat x$ and $\hat y$ are equivalent and we set $n_y = B_y = 0$. Second, Fig. \ref{totConcFig} shows that the concurrence $C(\tau)$ has its maximum for $B_z = 0$ and we thus set $B_z^{\rm ext} = -B_{\rm max}\eta n_z$. Finally we let $\tan\varphi = n_x/n_z$ and the total effective magnetic field is given by
\begin{equation}
\mathbf B = (\eta B_{\rm max}\sin\varphi + B_x^{\rm ext})\hat x,
\end{equation}
and depends on the three free parameters $\eta$, $\varphi$, and $B_x^{\rm ext}$.
For $\eta = 0.9$ the result is shown in Fig. \ref{concurrencePol1} and we find that for every $\varphi$ there are two applied magnetic fields along $B_x$ locally maximizing the concurrence. As expected from the discussion in the previous section, these occur when $B_x = \pm \Bcr$. We may thus set $B_x = \Bcr$ and study concurrence as a function of the polarization angle $\varphi$ which is shown in Fig. \ref{anglefig}, where we observe that the concurrence is maximized by minimizing fluctuations along $\hat z$.
Finally we can investigate the concurrence as a function of polarization, shown in the inset of Fig. \ref{anglefig}. We find that an increased nuclear spin polarization along $\hat z$ leads to an increased concurrence. We also see that a nuclear spin polarization perpendicular to $\hat z$ has almost no effect on the concurrence. This can be explained by the fact that the $g$-factors for the $x$- and $y$-directions are much smaller than the one along $z$.
\section{Summary}
We have theoretically investigated the entanglement between two photons emitted from a cascade recombination of a biexciton in a quantum dot. The entanglement was examined using the concurrence as a quantitative measure. We considered the two main sources of loss of concurrence, the FSS combined with a stochastic intermediate exciton lifetime and the stochastic Overhauser field. We found that the two sources of decoherence cannot be minimized independently of each other, and that the FSS is the dominant source of decoherence and must be minimized in order to maximize concurrence. Furthermore, we showed that reducing the uncertainty of the Overhauser field by nuclear spin polarization together with an applied magnetic field along a certain direction can improve the concurrence of the emitted light. The increase in entanglement depends strongly on the degree as well as the direction of nuclear spin polarization relative to the growth axis of the QD. This effect is caused by the difference between in-plane $g$-factors and the $g$-factor along the growth direction.
\section*{Acknowledgements}
We acknowledge funding from the Konstanz Center of Applied Photonics (CAP), DFG within SFB 767, and BMBF under the program Q.com-H.
\bibliographystyle{apsrev}

\clearpage
\newpage
\appendix
\section{The Probability Distribution of the Overhauser field in a QD}
\label{weightedaverage}
In this section we show that the probability distribution of a weighted sum of $N$ identical stochastic variables approaches a Gaussian distribution when $N \rightarrow \infty$. This is closely related to the well-known Central Limit Theorem\cite{athreya2006measure} in probability theory. Here we make small extension by considering a sum of identically distributed variables with different coefficients. The aim is to find the probability density function for the sum
\begin{equation}
Y = \sum_{n = 1}^Na_nX_n,
\end{equation}
where $X_n$ are identically distributed stochastic variables with expectation value $E[X]$ and variance $\sigma^2$, and $a_n$ are finite coefficients which, in general, also depend on $N$. For a QD we may choose the coefficients to match the electron position probability density function:
\begin{equation}
a_n = \frac{|\Psi\left(-l/2 + nl/N\right)|^2}{N},
\end{equation}
which implies that
\begin{align}
\begin{split}
\lim_{N\rightarrow\infty}\sum_{n=1}^N a_n &= \lim_{N\rightarrow\infty}\sum_{n=1}^N\frac{|\Psi\left(-l/2 + nl/N\right)|^2}{N}\\
= &\int_{-l/2}^{l/2}|\Psi\left(x\right)|^2\,dx = 1,
\end{split}
\end{align}
where $l$ is the size of the quantum dot. We introduce new variables with vanishing expectation value,
\begin{equation}
\tilde X_n = X_n-E[X_n],
\end{equation}
and form the new sum
\begin{equation}
\tilde Y = \sum_{n = 1}^Na_n\tilde X_n,
\end{equation}
which is related to $Y$ by
\begin{equation}
Y = \tilde Y + \Sigma(N)E[X],
\end{equation}
where
\begin{equation}
\Sigma(N) = \sum_{n=1}^{N}a_n.
\end{equation}
The characteristic function of $\tilde Y$ is given by
\begin{equation}
\varphi_{\tilde Y}(t) = \prod_{n = 1}^N \varphi_{\tilde X}\left(a_nt\right) \approx \prod_{n = 1}^N \left(1 - \frac{a_n^2\sigma^2t^2}{2}\right),
\end{equation}
where $\varphi_{\tilde X}$ is the characteristic function of any of the $\tilde X_n$ which we expand Taylor series to the second order in the second step. Now we consider
\begin{equation}
\varphi_{\tilde Y}(t)\approx \exp\left[\sum_{n = 1}^N\ln\left(1 - \frac{a_n^2\sigma^2t^2}{2}\right)\right]
\end{equation}
and
\begin{align}
&\sum_{n = 1}^N\ln\left(1 - \frac{a_n^2\sigma^2t^2}{2}\right) = -\sum_{n = 1}^N\sum_{k=1}^\infty\frac{a_n^{2k}\sigma^{2k}t^{2k}}{2^k}\\
\label{gaussianlimit}
\approx&-\frac{S_1\sigma^{2}t^{2}}{2N} -\sum_{k=2}^\infty\frac{S_k\sigma^{2k}t^{2k}}{2^kN^{2k-1}},
\end{align}
where
\begin{equation}
S_k = \int_0^l|\Psi(x)|^{4k}\,dx.
\end{equation}
We make the restriction that $|\Psi(x)|^2$ is bounded on $[0,l]$ which means that there is some constant $C$ such that $|\Psi(x)|^2 \leq C$. This also ensures the existence of all $S_k$ and for $N$ approaching infinity we may keep only the first term in Eq. (\ref{gaussianlimit}) and we find
\begin{equation}
\varphi_{\tilde Y}(t) \approx \exp\left(-\frac{S_1\sigma^{2}t^{2}}{2N}\right),
\end{equation}
and from this we obtain the probability density function of $\tilde Y$ as
\begin{equation}
f_{\tilde Y}(y) = \frac{\sqrt N}{\sigma\sqrt{2\pi S_1}}\exp\left(-\frac{Ny^2}{2S_1\sigma^2}\right),
\end{equation}
which is a Gaussian distribution with variance $S_1\sigma^2/N$. $S_1$ depends on the the coefficients $a_n$ but the general form is always a Gaussian distribution regardless of what wave function is considered.
\end{document}